\documentclass[12pt]{article}
\usepackage{oldgerm}
\usepackage{yfonts}
\usepackage{euler}
\usepackage{amsmath}
\usepackage{amssymb}
\usepackage{amstext}
\usepackage{amscd}
\usepackage{amsfonts}
\DeclareMathAlphabet{\EuFrak}{U}{euf}{m}{n}
\DeclareMathAlphabet{\EuScript}{U}{eus}{m}{n}

\title{{\bf Convolution of n-dimensional Tempered Ultradistributions and
Field Theory}
\thanks{\it{This work was partially supported by Consejo Nacional de
Investigaciones Cient\'{\i}ficas and Comisi\'{o}n de
Investigaciones Cient\'{\i}ficas de la Pcia. de Buenos Aires;
Argentina.}}}
\author{C.G.Bollini and M.C.Rocca\\
Departamento de F\'{\i}sica, Fac. de Ciencias Exactas,\\
Universidad Nacional de La Plata.\\
C.C. 67 (1900) La Plata. Argentina.}

\date{April 1, 2002}

\begin{document}

\maketitle

\vspace{-5mm}

\begin{abstract}

In this work, a general definition of convolution between two
arbitrary Tempered Ultradistributions is given. When one of the
Tempered Ultradistributions is rapidly decreasing this definition
coincides with the definition of J. Sebastiao e Silva. In the
four-dimensional case, when the Tempered Ultradistributions are
even in the variables $k^0$ and $\rho$ (see Section 5) we obtain
an expression for the convolution, which is more suitable for
practical applications. The product of two arbitrary even (in the
variables $x^0$ and $r$) four dimensional distributions of
exponential type is defined via the convolution of its
corresponding Fourier Transforms. With this definition of
convolution , we treat the problem of singular products of Green
Functions in Quantum Field Theory. (For Renormalizable as well as
for Nonrenormalizable Theories). Several examples of convolution
of two Tempered Ultradistributions are given. In particular we
calculate the convolution of two massless Wheeeler's propagators
and the convolution of two complex mass Wheeler's propagators.


\end{abstract}

\newpage

\renewcommand{\theequation}{\arabic{section}.\arabic{equation}}

\section{Introduction}

The question of the product of distributions with coincident point
singularities is related in Field Theory, to the asymptotic
behavior of loop integrals of propagators.

From a mathematical point of view, practically all definitions lead
to limitations on the set of distributions that can be multiplied
together to give another distribution of the same kind.

The properties of ultradistributions (ref.\cite{tp1,tp2}) are well
adapted for their use in Field Theory. In this respect we have
shown (ref.\cite{tp3}) that it is possible to define in one
dimensional space, the convolution of any pair of tempered
ultradistributions, giving as a result another tempered
ultradistribution. The next step is to consider the convolution
of any pair of tempered ultradistribution in n-dimensional space.
As we shall see, this follows from the formula obtained
in ref.\cite{tp3} for one dimensional space.

However, the resultant formula is rather complex  to be used
in practical applications and calculus. Then, for applications,
it is convenient to consider the convolution of any two
tempered ultradistributions which are even in the variables $k^0$ y $\rho$
(see section 5).

Ultradistributions also have the
advantage of being representable by means of analytic functions.
So that, in general, they are easier to work with them and,
as we shall see, have interesting properties. One of those properties
is that Schwartz tempered distributions are canonical and continuously
injected
into tempered ultradistributions and as a consequence the Rigged
Hilbert Space with tempered distributions is canonical and continuously
included
in the Rigged Hilbert Space with tempered ultradistributions.

This paper is organized as follow:
in sections 2 and 3 we define the Distributions of Exponential Type
and the Fourier transformed Tempered Ultradistributions. Each of them
is part of a Guelfand's Triplet ( or Rigged Hilbert Space \cite{tp4} )
together with their respective duals and a ``middle term'' Hilbert
space. In section 4 we give a general expression for the convolution
of any pair of n-dimensional tempered ultradistributions and some
simple examples. In section 5 we obtain  the expression for the
convolution of any pair of even tempered ultradistributions.
In section 6, we evaluate the convolution of
two massless Wheeler's propagators. In section 7 we evaluate the
convolution of two complex mass Wheeler's propagators.
Finally, section 8 is reserved for a discussion of the principal results.
For the benefit of the reader an Appendix is added containing some
formulas utilized in the text.

\section{Distributions of Exponential Type}

For the sake of the reader we shall present a brief description of the
principal properties of Tempered Ultradistributions.

{\bf Notations}.
The notations are almost textually taken from ref\cite{tp2}.
Let $\boldsymbol{{\mathbb{R}}^n}$
(res. $\boldsymbol{{\mathbb{C}}^n}$) be the real (resp. complex)
n-dimensional space whose points are denoted by $x=(x_1,x_2,...,x_n)$
(resp $z=(z_1,z_2,...,z_n)$). We shall use the notations:

(i) $x+y=(x_1+y_1,x_2+y_2,...,x_n+y_n)$\; ; \;
    $\alpha x=(\alpha x_1,\alpha x_2,...,\alpha x_n)$

(ii)$x\geqq 0$ means $x_1\geqq 0, x_2\geqq 0,...,x_n\geqq 0$

(iii)$x\cdot y=\sum\limits_{j=1}^n x_j y_j$

(iV)$\mid x\mid =\sum\limits_{j=1}^n \mid x_j\mid$

Let $\boldsymbol{{\mathbb{N}}^n}$ be the set of n-tuples of natural
numbers. If $p\in\boldsymbol{{\mathbb{N}}^n}$, then
$p=(p_1, p_2,...,p_n)$,
and $p_j$ is a natural number, $1\leqq j\leqq n$. $p+q$ denote
$(p_1+q_1, p_2+q_2,..., p_n+q_n)$ and $p\geqq q$ means $p_1\geqq q_1,
p_2\geqq q_2,...,p_n\geqq q_n$. $x^p$ means $x_1^{p_1}x_2^{p_2}...
x_n^{p_n}$. We shall denote by
$\mid p\mid=\sum\limits_{j=1}^n  p_j $ and by $D^p$ we denote the
differential operator ${\partial}^{p_1+p_2+...+p_n}/\partial{x_1}^{p_1}
\partial{x_2}^{p_2}...\partial{x_n}^{p_n}$

For any natural $k$ we define $x^k=x_1^k x_2^k...x_n^k$
and ${\partial}^k/\partial x^k=
{\partial}^{nk}/\partial x_1^k\partial x_2^k...\partial x_n^k$

The space $\boldsymbol{{\cal H}}$  of test functions
such that $e^{p|x|}|D^q\phi(x)|$ is bounded for any p and q
is defined ( ref.\cite{tp2} ) by means
of the countably set of norms:
\begin{equation}
\label{ep2.1}
{\|\hat{\phi}\|}_p=\sup_{0\leq q\leq p,\,x}
e^{p|x|} \left|D^q \hat{\phi} (x)\right|\;\;\;,\;\;\;p=0,1,2,...
\end{equation}
According to reference\cite{tp5} $\boldsymbol{{\cal H}}$  is a
$\boldsymbol{{\cal K}\{M_p\}}$ space
with:
\begin{equation}
\label{ep2.2}
M_p(x)=e^{(p-1)|x|}\;\;\;,\;\;\; p=1,2,...
\end{equation}
$\boldsymbol{{\cal K}\{e^{(p-1)|x|}\}}$ satisfies condition
$\boldsymbol({\cal N})$
of Guelfand ( ref.\cite{tp4} ). It is a countable Hilbert and nuclear
space:
\begin{equation}
\label{ep2.3}
\boldsymbol{{\cal K}\{e^{(p-1)|x|}\}} =\boldsymbol{{\cal H}} =
\bigcap\limits_{p=1}^{\infty}\boldsymbol{{\cal H}_p}
\end{equation}
where $\boldsymbol{{\cal H}_p}$ is obtained by completing
$\boldsymbol{{\cal H}}$ with the norm induced by
the scalar product:
\begin{equation}
\label{ep2.4}
{<\hat{\phi}, \hat{\psi}>}_p = \int\limits_{-\infty}^{\infty}
e^{2(p-1)|x|} \sum\limits_{q=0}^p D^q \overline{\hat{\phi}} (x) D^q
\hat{\psi} (x)\;dx \;\;\;;\;\;\;p=1,2,...
\end{equation}
where $dx=dx_1\;dx_2...dx_n$

If we take the usual scalar product:
\begin{equation}
\label{ep2.5}
<\hat{\phi}, \hat{\psi}> = \int\limits_{-\infty}^{\infty}
\overline{\hat{\phi}}(x) \hat{\psi}(x)\;dx
\end{equation}
then $\boldsymbol{{\cal H}}$, completed with (2.5), is the Hilbert space
$\boldsymbol{H}$
of square integrable functions.

The space of continuous linear functionals defined on
$\boldsymbol{{\cal H}}$ is the space
$\boldsymbol{{\Lambda}_{\infty}}$ of the distributions of the exponential
type ( ref.\cite{tp2} ).

The ``nested space''
\begin{equation}
\label{ep2.6}
\textgoth{\Large{H}}=
\boldsymbol{(}\boldsymbol{{\cal H}},\boldsymbol{H},
\boldsymbol{{\Lambda}_{\infty}} \boldsymbol{)}
\end{equation}
is a Guelfand's triplet ( or a Rigged Hilbert space \cite{tp4} ).

In addition we have: $\boldsymbol{{\cal H}}\subset\boldsymbol{{\cal S}}
\subset\boldsymbol{H}\subset\boldsymbol{{\cal S}^{'}}\subset
\boldsymbol{{\Lambda}_{\infty}}$, where $\boldsymbol{{\cal S}}$ is the
Schwartz space of rapidly decreasing test functions (ref\cite{tp6}).

Any Guelfand's triplet
$\textgoth{\Large{G}}=\boldsymbol{(}\boldsymbol{\Phi},
\boldsymbol{H},\boldsymbol{{\Phi}^{'}}\boldsymbol{)}$
has the fundamental property that a linear and symmetric operator
on $\boldsymbol{\Phi}$, admitting an extension to a self-adjoint
operator in
$\boldsymbol{H}$, has a complete set of generalized eigen-functions
in $\boldsymbol{{\Phi}^{'}}$ with real eigenvalues.

\section{Tempered Ultradistributions}
\setcounter{equation}{0}

The Fourier transform of a function $\hat{\phi}\in \boldsymbol{{\cal H}}$
is
\begin{equation}
\label{ep3.1}
\phi(z)=\frac {1} {2\pi}
\int\limits_{-\infty}^{\infty}\overline{\hat{\phi}}(x)\;e^{iz\cdot x}\;dx
\end{equation}
$\phi(z)$ is entire analytic and rapidly decreasing on straight lines
parallel
to the real axis. We shall call $\boldsymbol{{\EuFrak H}}$
the set of all such functions.
\begin{equation}
\label{ep3.2}
\boldsymbol{{\EuFrak H}}={\cal F}\left\{\boldsymbol{{\cal H}}\right\}
\end{equation}
It is a $\boldsymbol{{\cal Z}\{M_p\}}$ space ( ref.\cite{tp5} ),
countably normed and complete, with:
\begin{equation}
\label{ep3.3}
M_p(z)= (1+|z|)^p
\end{equation}
$\boldsymbol{{\EuFrak H}}$ is also a nuclear space with norms:
\begin{equation}
\label{ep3.4}
{\|\phi\|}_{pn} = \sup_{z\in V_n} {\left(1+|z|\right)}^p
|\phi (z)|
\end{equation}
where $V_k=\{z=(z_1,z_2,...,z_n)\in\boldsymbol{{\mathbb{C}}^n}:
\mid Im z_j\mid\leqq k, 1\leqq j \leqq n\}$

We can define the usual scalar product:
\begin{equation}
\label{ep3.5}
<\phi (z), \psi (z)>=\int\limits_{-\infty}^{\infty}
\phi(z) {\psi}_1(z)\;dz =
\int\limits_{-\infty}^{\infty} \overline{\hat{\phi}}(x)
\hat{\psi}(x)\;dx
\end{equation}
where:
\[{\psi}_1(z)=\int\limits_{-\infty}^{\infty}
\hat{\psi}(x)\; e^{-iz\cdot x}\;dx\]
and $dz=dz_1\;dz_2...dz_n$

By completing $\boldsymbol{{\EuFrak H}}$ with the norm induced by (3.5)
we get the Hilbert space of square integrable functions.

The dual of $\boldsymbol{{\EuFrak H}}$ is the space
$\boldsymbol{{\cal U}}$ of tempered ultradistributions
( ref.\cite{tp2} ). In other words, a tempered ultradistribution is
a continuous linear functional defined on the space
$\boldsymbol{{\EuFrak H}}$ of entire
functions rapidly decreasing on straight lines parallel to the real axis.

The set
$\textgoth{\Large{U}}=
\boldsymbol{({\EuFrak H},H,{\cal U})}$ is also a Guelfand's triplet.

Moreover, we have: $\boldsymbol{{\EuFrak H}}\subset\boldsymbol{{\cal S}}
\subset\boldsymbol{H}\subset\boldsymbol{{\cal S}^{'}}\subset
\boldsymbol{{\cal U}}$.

$\boldsymbol{{\cal U}}$ can also be characterized in the following way
( ref.\cite{tp2} ): let $\boldsymbol{{\cal A}_{\omega}}$ be the space of
all functions $F(z)$ such that:

${\Large {\boldsymbol{I}}}$-
$F(z)$ is analytic for $\{z\in \boldsymbol{{\mathbb{C}}^n} :
|Im(z_1)|>p, |Im(z_2)|>p,...,|Im(z_n)|>p\}$.

${\Large {\boldsymbol{II}}}$-
$F(z)/z^p$ is bounded continuous  in
$\{z\in \boldsymbol{{\mathbb{C}}^n} :|Im(z_1)|\geqq p,|Im(z_2)|\geqq p,
...,|Im(z_n)|\geqq p\}$,
where $p=0,1,2,...$ depends on $F(z)$.

Let $\boldsymbol{\Pi}$ be the set of all $z$-dependent pseudo-polynomials,
$z\in \boldsymbol{{\mathbb{C}}^n}$.
Then $\boldsymbol{{\cal U}}$ is the quotient space:

${\Large {\boldsymbol{III}}}$-
$\boldsymbol{{\cal U}}=\boldsymbol{{\cal A}_{\omega}/\Pi}$

By a pseudo-polynomial we understand a function of $z$ of the form $\;\;$
$\sum_s z_j^s G(z_1,...,z_{j-1},z_{j+1},...,z_n)$ with
$G(z_1,...,z_{j-1},z_{j+1},...,z_n)\in\boldsymbol{{\cal A}_{\omega}}$

Due to these properties it is possible to represent any ultradistribution
as ( ref.\cite{tp2} ):
\begin{equation}
\label{ep3.6}
F(\phi)=<F(z), \phi(z)>=\oint\limits_{\Gamma} F(z) \phi(z)\;dz
\end{equation}
$\Gamma={\Gamma}_1\cup{\Gamma}_2\cup ...{\Gamma}_n$
where the path ${\Gamma}_j$ runs parallel to the real axis from
$-\infty$ to $\infty$ for $Im(z_j)>\zeta$, $\zeta>p$ and back from
$\infty$ to $-\infty$ for $Im(z_j)<-\zeta$, $-\zeta<-p$.
( $\Gamma$ surrounds all the singularities of $F(z)$ ).

Formula (3.6) will be our fundamental representation for a tempered
ultradistribution. Sometimes use will be made of ``Dirac formula''
for ultradistributions ( ref.\cite{tp1} ):
\begin{equation}
\label{ep3.7}
F(z)=\frac {1} {(2\pi i)^n}\int\limits_{-\infty}^{\infty}
\frac {f(t)} {(t_1-z_1)(t_2-z_2)...(t_n-z_n)}\;dt
\end{equation}
where the ``density'' $f(t)$ is such that
\begin{equation}
\label{ep3.8}
\oint\limits_{\Gamma} F(z) \phi(z)\;dz =
\int\limits_{-\infty}^{\infty} f(t) \phi(t)\;dt
\end{equation}
While $F(z)$ is analytic on $\Gamma$, the density $f(t)$ is in
general singular, so that the r.h.s. of (3.8) should be interpreted
in the sense of distribution theory.

Another important property of the analytic representation is the fact
that on $\Gamma$, $F(z)$ is bounded by a power of $z$ ( ref.\cite{tp2} ):
\begin{equation}
\label{ep3.9}
|F(z)|\leq C|z|^p
\end{equation}
where $C$ and $p$ depend on $F$.

The representation (3.6) implies that the addition of a
pseudo-polynomial $P(z)$ to $F(z)$ do not alter the ultradistribution:
\[\oint\limits_{\Gamma}\{F(z)+P(z)\}\phi(z)\;dz=
\oint\limits_{\Gamma} F(z)\phi(z)\;dz+\oint\limits_{\Gamma}
P(z)\phi(z)\;dz\]
But:
\[\oint\limits_{\Gamma} P(z)\phi(z)\;dz=0\]
as $P(z)\phi(z)$ is entire analytic in some of the variables $z_j$
( and rapidly decreasing ),
\begin{equation}
\label{ep3.10}
\therefore \;\;\;\;\oint\limits_{\Gamma} \{F(z)+P(z)\}\phi(z)\;dz=
\oint\limits_{\Gamma} F(z)\phi(z)\;dz
\end{equation}

\section{The Convolution}

\setcounter{equation}{0}

In ref.\cite{tp3} we have defined and shown the existence of the
convolution product between to arbitrary one dimensional tempered
ultradistributions.

We now define:
\begin{equation}
\label{ep4.1}
H_{\lambda}(k)=\frac {i} {(2\pi)^n}
\oint\limits_{{\Gamma}_1}
\oint\limits_{{\Gamma}_2}
\frac {k_1^{\lambda} F(k_1) k_2^{\lambda} G(k_2)}
{k-k_1-k_2}\;dk_1\;dk_2
\end{equation}
$(k-k_1-k_2={\prod}_{i=1}^n (k_i-k_{1i}-k_{2i}))$.
Let ${\textgoth{b}}_i$ be a vertical band contained in the
${\lambda}_i$-plane ${\textgoth{p}}_i$
Integral (\ref{ep4.1}) is an analytic function of $\lambda$ defined in a
domain \textgoth{B} given by the cartesian product of vertical  bands
$\prod {\textgoth{b}}_i $  contained
in  the cartesian product $\textgoth{P}=\prod {\textgoth{p}}_i$
of the n $\lambda$-planes. Moreover, it is bounded by a power of
$|k|$.
Then, according to the method of ref.\cite{tp7}, $H_{\lambda}$ can be
analytically continued to other parts of \textgoth{P}.
In particular near the origin we have the Laurent
expansion:
\begin{equation}
\label{ep4.2}
H_{\lambda}(k)=\sum\limits_nH^{(n)}(k){\lambda}^n
\end{equation}
We now define
the convolution product as the $\lambda$-independent term of
(\ref{ep4.2}):
\begin{equation}
\label{ep4.3}
H(k)=H^{(0)}(k)
\end{equation}
The proof that $H^{(0)}(k)$ is a Tempered Ultradistribution is similar
to the one given in ref.\cite{tp3} for the one-dimensional case.
 For an immediate application of (\ref{ep4.1}-\ref{ep4.3}) we can
 evaluate
 the product of two arbitrary derivatives of a n-dimensional $\delta$
 distribution.
 By calculating the convolution product of
 the Fourier transforms of ${\delta}^{(m)}(x)$ an ${\delta}^{(n)}(x)$,
and then antitransforming, we can show that:
\begin{equation}
\label{ep4.4}
{\delta}^{(m)}(x)\cdot {\delta}^{(n)}(x)=0
\end{equation}
extending the result obtained in ref.\cite{tp3} for the one-dimensional
case.

Likewise, we can obtain:
\begin{equation}
\label{ep4.5}
(x_{1+}^{{\alpha}_1}\ x_{2+}^{{\alpha}_2}... x_{n+}^{{\alpha}_n})\cdot
(x_{1+}^{{\beta}_1}\ x_{2+}^{{\beta}_2}... x_{n+}^{{\beta}_n})=
(x_{1+}^{{\alpha}_1+{\beta}_1}\ x_{2+}^{{\alpha}_2+{\beta}_2}...
x_{n+}^{{\alpha}_n+{\beta}_n})
\end{equation}
generalizing again the result of ref.\cite{tp3}.

As another example let us consider the product
$(x^{-n_1}y^{-m_1})\cdot(x^{-n_2}y^{-m_2})$

We have
\[{\cal F}\{(x^{-n_1}y^{-m_1})\cdot(x^{-n_2}y^{-m_2})\}=\frac {(-i)^{n_1+n_2}}
{(n_1+n_2-1)!}z_1^{n_1+n_2-1}\left[\frac {i} {4}\frac {z_1^{2{\lambda}_1}} {{\lambda}_1}
+ \frac {i} {2} \ln(z_1)+\right.\]
\[\left.\frac {\pi} {2} Sgn[\Im(z_1)]\right]
\frac {(-i)^{m_1+m_2}}
{(m_1+m_2-1)!}z_2^{m_1+m_2-1}\left[\frac {i} {4}\frac {z_2^{2{\lambda}_2}} {{\lambda}_2}
+ \frac {i} {2} \ln(z_2)+\frac {\pi} {2} Sgn[\Im(z_2)]\right]=\]
\[\frac {(-i)^{n_1+n_2}} {(n_1+n_2-1)!} z_1^{n_1+n_2-1}
\left[\frac {i} {4{\lambda}_1}\left[1+2{\lambda}_1\ln(z_1)\right]
+ \frac {i} {2} \ln(z_1)+ \frac {\pi} {2} Sgn[\Im(z_1)]\right]\times\]
\begin{equation}
\label{ep4.6}
\hspace{-7.5mm} \frac {(-i)^{m_1+m_2}}
{(m_1+m_2-1)!}z_2^{m_1+m_2-1}\left[\frac {i} {4{\lambda}_2}\left[1+2{\lambda}_2\ln(z_2)
\right]
+ \frac {i} {2} \ln(z_2)+\frac {\pi} {2} Sgn[\Im(z_2)]\right]
\end{equation}
The $({\lambda}_1;{\lambda}_2)$-independent term is:
\[\frac {(-i)^{n_1+n_2}\pi} {(n_1+n_2-1)!} z_1^{n_1+n_2-1}
\left[\frac {1} {\pi i}\ln(z_1)
-\frac {\pi} {2} Sgn[\Im(z_1)]\right]\times\]
\begin{equation}
\label{ep4.7}
\hspace{-7.5mm} \frac {(-i)^{m_1+m_2}}
{(m_1+m_2-1)!}z_2^{m_1+m_2-1}\left[\frac {1} {\pi i} \ln(z_2)-
\frac {\pi} {2} Sgn[\Im(z_2)]\right]
\end{equation}
and it is recognized to be ${\cal F}\{x^{-n_1-n_2}y^{-m_1-m_2}\}$

\section{The Convolution of even four-dimensional Tempered
Ultradistributions}

\setcounter{equation}{0}

We pass now to consider the convolution of two even tempered
ultradistributions.

The Fourier transform of a distribution of exponential
type, even in the variables $x^0$ and $|\vec{x}|$
is by definition a even tempered ultradistribution in the variables
$k^0$ and $\rho=(k_1^2+k_2^2+\cdot\cdot\cdot+k_n^2)^{1/2}$
Taking into account the equality:
\begin{equation}
\label{ep5.1}
\int\limits_{-\infty}^{+\infty}\hat{f}(x)\hat{\phi}(x)\;dx=
\oint\limits_{\Gamma}F(k)\phi(k)\;dk=
\int\limits_{-\infty}^{+\infty}f(k)\phi(k)\;dk
\end{equation}
(where $F(k)$ and $f(k)$ are related by (\ref{ep3.7})) we conclude that
$f(k)$ is even in $k^0$ and $\rho$.

For most practical applications one has to deal with the convolution of two
Lorentz invariant ultradistributions. They are particular cases of ultradistributions
which are even in two relevant variables: one temporal and the other
the spacial distance (The even ultradistributions).

Let as now consider  $\hat{f}\in \boldsymbol{H}$ even. Then we can write:
\begin{equation}
\label{ep5.2}
\hat{f}(x_0,r)=\frac {i} {(2\pi)^3r}\iint\limits_{-\infty}^{+\infty}
f(k_0,\rho) e^{-ik^0x^0}e^{-i\rho r}\;\rho\;d\rho
dk^0
\end{equation}
\begin{equation}
\label{ep5.3}
f(k_0,\rho)=-\frac {2\pi i} {\rho}\iint\limits_{-\infty}^{+\infty}
\hat{f}(x_0,r) e^{ik^0x^0}e^{i\rho r}\;r\;dr
dx^0
\end{equation}
Let as now take $\hat{g}\in\boldsymbol{H}$. Then according to (\ref{ep5.2}):
\[\hat{f}(x)\hat{g}(x)=-\frac {1} {(2\pi)^6r^2}
\iiiint\limits_{-\infty}^{+\infty}f(k^0_1,{\rho}_1)g(k^0_2,{\rho}_2)
e^{-i(k^0_1+k^0_2)x^0}e^{-i({\rho}_1+{\rho}_2)r}\;\times\]
\begin{equation}
\label{ep5.4}
\times\;{\rho}_1{\rho}_2\;d{\rho}_1\;d{\rho}_2\;dk^0_1\;dk^0_2
\end{equation}
and Fourier transforming (\ref{ep5.4})
\[\boldsymbol{{\cal F}}\{\hat{f}(x)\hat{g}(x)\}(k)=\frac {i}
{(2\pi)^5\rho}\idotsint\limits_{-\infty}^{+\infty}
f(k^0_1,{\rho}_1)g(k^0_2,{\rho}_2)
e^{i(k^0-k^0_1-k^0_2)x^0}e^{i(\rho-{\rho}_1-{\rho}_2)r}\;\times\]
\begin{equation}
\label{ep5.5}
\times\;{\rho}_1{\rho}_2\;d{\rho}_1\;d{\rho}_2\;dk^0_1\;dk^0_2\;
r^{-1}dr\;dx^0
\end{equation}
Evaluating the integral in the variable $x^0$ and calling
$h(k^0,\rho)=\boldsymbol{{\cal F}}\{\hat{f}(x)\hat{g}(x)\}(k)$
in (\ref{ep5.5}) we obtain
\[h(k^0,\rho)=i\idotsint\limits_{-\infty}^{+\infty}
f(k^0_1,{\rho}_1)g(k^0_2,{\rho}_2)
\delta (k^0-k^0_1-k^0_2)
\frac {e^{i(\rho-{\rho}_1-{\rho}_2)r}} {\rho}\;\times\]
\begin{equation}
\label{ep5.6}
\times\;{\rho}_1{\rho}_2\;d{\rho}_1\;d{\rho}_2\;dk^0_1\;dk^0_2\;
r^{-1}dr
\end{equation}
We want now to extend $h(k^0,\rho)$ to the complex plane as a
tempered ultradistribution. For this we can use for example,
formula (\ref{ep3.7}). First we consider the term
\begin{equation}
\label{ep5.7}
\frac {e^{i(\rho-{\rho}_1-{\rho}_2)r}} {\rho}
\end{equation}
The extension to the complex plane is:
\begin{equation}
\label{ep5.8}
\{\Theta(r)\;\Theta[\Im(\rho)]-\Theta(-r)\;\Theta[-\Im(\rho)]\}
\;\frac {e^{i(\rho-{\rho}_1-{\rho}_2)r}} {\rho}
\end{equation}
where $\Theta$ is the Heaviside's step function and $\Im$ denotes
``Imaginary part''.

On the other hand the extension of
\begin{equation}
\label{ep5.9}
\delta(k^0-k^0_1-k^0_2)
\end{equation}
is
\begin{equation}
\label{ep5.10}
-\frac {1} {2\pi i(k^0-k^0_1-k^0_2)}
\end{equation}
Replacing [(\ref{ep5.8}),(\ref{ep5.10})] in (\ref{ep5.6}) and
then integrating out the variable $r$ we obtain:
\[H(k^0,\rho)=\frac {1} {2\pi\rho}\iiiint\limits_{-\infty}^{+\infty}
\frac {f(k^0_1,{\rho}_1)g(k^0_2,{\rho}_2)} {k^0-k^0_1-k^0_2}
\{\Theta[\Im(\rho)]\ln ({\rho}_1+{\rho}_2-\rho)+\Theta[-\Im(\rho)]
\;\times\]
\begin{equation}
\label{ep5.11}
\ln(\rho-{\rho}_1-{\rho}_2)\}\;
{\rho}_1{\rho}_2\;d{\rho}_1\;d{\rho}_2\;dk^0_1\;dk^0_2
\end{equation}
where $H(k^0,\rho)$ is the extension of $f(k^0,\rho)$.
Taking into account that $f(k^0_1,{\rho}_1)$ and $g(k^0_2,{\rho}_2)$
are even functions in the first and second variables (\ref{ep5.11}) takes
the
form:
\[H(k^0,\rho)=\frac {1} {4\pi\rho}\iiiint\limits_{-\infty}^{+\infty}
\frac {f(k^0_1,{\rho}_1)g(k^0_2,{\rho}_2)} {k^0-k^0_1-k^0_2}
\ln [{\rho}^2-({\rho}_1+{\rho}_2)^2]
\;\times\]
\begin{equation}
\label{ep5.12}
{\rho}_1{\rho}_2\;d{\rho}_1\;d{\rho}_2\;dk^0_1\;dk^0_2
\end{equation}
The expression (\ref{ep5.12}) for  $H(k^0,\rho)$ can be re-writted in the
form
\[H(k^0,\rho)=\frac {1} {4\pi\rho}\oint\limits_{{\Gamma}^0_1}
\oint\limits_{{\Gamma}^0_2}\oint\limits_{{\Gamma}_1}
\oint\limits_{{\Gamma}_2}
\frac {F(k^0_1,{\rho}_1)G(k^0_2,{\rho}_2)} {k^0-k^0_1-k^0_2}
\ln [{\rho}^2-({\rho}_1+{\rho}_2)^2]\;\times\]
\begin{equation}
\label{ep5.13}
{\rho}_1{\rho}_2\;d{\rho}_1\;d{\rho}_2\;dk^0_1\;dk^0_2
\end{equation}
where $F(k^0_1,{\rho}_1)$ and $G(k^0_2,{\rho}_2)$ are respectively, the
extensions of $f(k^0_1,{\rho}_1)$ and $g(k^0_2,{\rho}_2)$ and where we
have taken: $|\Im(k^0)|>|\Im(k^0_1)|+|\Im(k^0_2)|$, $|\Im(\rho)|>$
$|\Im({\rho}_1)|+|\Im({\rho}_2)|$. In addition ${\Gamma}^0_1$,
${\Gamma}^0_2$,${\Gamma}_1$ and ${\Gamma}_2$ are respectively, paths
(as we have described in section 3 ), in the variables
$k^0_1, k^0_2,{\rho}_1$
and ${\rho}_2$, enclosing all the singularities of the integrand in
(\ref{ep5.13}). The difference between
\[\int\frac {2\rho} {{\rho}^2-({\rho}_1+{\rho}_2)^2}\;d\rho\;\;\;\;\;\;
\mathrm{and}\;\;\;\;\;\;\ln[{\rho}^2-({\rho}_1+{\rho}_2)^2]\]
is an entire analytic function. With this substitution in
(\ref{ep5.13}) we
obtain
\[H(k^0,\rho)=\frac {1} {2\pi\rho}
\int\rho\;d\rho
\oint\limits_{{\Gamma}^0_1}
\oint\limits_{{\Gamma}^0_2}\oint\limits_{{\Gamma}_1}
\oint\limits_{{\Gamma}_2}
\frac {F(k^0_1,{\rho}_1)G(k^0_2,{\rho}_2)} {k^0-k^0_1-k^0_2}
\frac {1} {{\rho}^2-({\rho}_1+{\rho}_2)^2}\;\times\]
\begin{equation}
\label{ep5.14}
{\rho}_1{\rho}_2\;d{\rho}_1\;d{\rho}_2\;dk^0_1\;dk^0_2
\end{equation}
Now we can use the method of ref.\cite{tp3} to define the convolution
for the case in which $F(k^0_1,{\rho}_1)$ and $G(k^0_2,{\rho}_2)$ are
tempered ultradistributions. We define:
\[H_{{\lambda}_0\lambda}(k^0,\rho)=\frac {1} {2\pi\rho}
\int\rho\;d\rho
\oint\limits_{{\Gamma}^0_1}
\oint\limits_{{\Gamma}^0_2}\oint\limits_{{\Gamma}_1}
\oint\limits_{{\Gamma}_2}
\frac {k^{0\;{\lambda}_0}_1{\rho}^{{\lambda}+1}_1F(k^0_1,{\rho}_1)
k^{0\;{\lambda}_0}_2{\rho}^{{\lambda}+1}_2G(k^0_2,{\rho}_2)}
{k^0-k^0_1-k^0_2} \times\]
\begin{equation}
\label{ep5.15}
\frac {1} {{\rho}^2-({\rho}_1+{\rho}_2)^2}\;
d{\rho}_1\;d{\rho}_2\;dk^0_1\;dk^0_2
\end{equation}
Integral (\ref{ep5.15}) is an analytic function of $({\lambda}_0,\lambda)$
bounded by a power of $|k|$ and defined
in a domain \textgoth{B} given by  the cartesian product
of a vertical band ${\textgoth{b}}_0$ contained in the
${\lambda}_0$-plane and vertical band \textgoth{b}
contained in the $\lambda$-plane.
We can again extend this domain using the method given in
ref.\cite{tp7}
and perform the Laurent expansion :
\begin{equation}
\label{ep5.16}
H_{{\lambda}_0\lambda}(k^0,\rho)=
\sum\limits_{mn} H^{(m,n)}(k^0,\rho){\lambda}^{m}_0{\lambda}^n
\end{equation}
We define the convolution product as the $({\lambda}_0,\lambda)$-
independent term of (\ref{ep5.16}).
\begin{equation}
\label{ep5.17}
H(k)=H(k^0,\rho)=H^{(0,0)}(k^0,\rho)
\end{equation}
The proof that $H(k)$ is an ultradistribution is similar
to the one given in ref.\cite{tp3}
for the one-dimensional case.

To simplify the evaluation of (\ref{ep5.15}) we define:
\[L_{{\lambda}_0\lambda}(k^0,\rho)=
\oint\limits_{{\Gamma}^0_1}
\oint\limits_{{\Gamma}^0_2}\oint\limits_{{\Gamma}_1}
\oint\limits_{{\Gamma}_2}
\frac {k^{0\;{\lambda}_0}_1{\rho}^{{\lambda}+1}_1F(k^0_1,{\rho}_1)
k^{0\;{\lambda}_0}_2{\rho}^{{\lambda}+1}_2G(k^0_2,{\rho}_2)}
{k^0-k^0_1-k^0_2} \times\]
\begin{equation}
\label{ep5.18}
\frac {1} {{\rho}^2-({\rho}_1+{\rho}_2)^2}\;
d{\rho}_1\;d{\rho}_2\;dk^0_1\;dk^0_2
\end{equation}
so that
\begin{equation}
\label{ep5.19}
H_{{\lambda}_0\lambda}(k^0,\rho)=\frac {1} {2\pi\rho}\int
L_{{\lambda}_0\lambda}(k^0,\rho)\;\rho\;d\rho
\end{equation}
Now we go to show that the cut on the real axis of
(\ref{ep5.17}) $h_{{\lambda}_0\lambda}(k^0,\rho)$
 is a even function of $k^0$ and  $\rho$.
For this purpose we consider
\[H_{{\lambda}_0\lambda}(k^0,\rho)=\frac {1} {4\pi\rho}
\oint\limits_{{\Gamma}^0_1}
\oint\limits_{{\Gamma}^0_2}\oint\limits_{{\Gamma}_1}
\oint\limits_{{\Gamma}_2}
\frac {k^{0\;{\lambda}_0}_1{\rho}^{{\lambda}+1}_1F(k^0_1,{\rho}_1)
k^{0\;{\lambda}_0}_2{\rho}^{{\lambda}+1}_2G(k^0_2,{\rho}_2)}
{k^0-k^0_1-k^0_2} \times\]
\begin{equation}
\label{ep5.20}
\ln[{\rho}^2-({\rho}_1+{\rho}_2)^2]\;
d{\rho}_1\;d{\rho}_2\;dk^0_1\;dk^0_2
\end{equation}
(\ref{ep5.20}) is explicitly odd in $\rho$. For the variable $k^0$
we take on account that $e^{i\pi{\lambda}_0\{Sgn[\Im(k^0_1)]+
Sgn[\Im(k^0_2)]\}}=1$ and as a consequence (\ref{ep5.20}) is odd
in $k^0$ too. We consider now the parity in variable $\rho$.
\[\oint\limits_{{\Gamma}_0}\oint\limits_{\Gamma}
H_{{\lambda}_0\lambda}(k^0,-\rho)\phi(k^0,\rho)\;dk^0\;d{\rho}=
-\iint\limits_{-\infty}^{+\infty}
h_{{\lambda}_0\lambda}(k^0,-\rho)\phi(k^0,\rho)\;dk^0\;d{\rho}=\]
\begin{equation}
\label{ep5.21}
-\oint\limits_{{\Gamma}_0}\oint\limits_{\Gamma}
H_{{\lambda}_0\lambda}(k^0,\rho)\phi(k^0,\rho)\;dk^0\;d{\rho}=
-\iint\limits_{-\infty}^{+\infty}
h_{{\lambda}_0\lambda}(k^0,\rho)\phi(k^0,\rho)\;dk^0\;d{\rho}
\end{equation}
Thus we have
\begin{equation}
\label{ep5.22}
h_{{\lambda}_0\lambda}(k^0,-\rho)=h_{{\lambda}_0\lambda}(k^0,\rho)
\end{equation}
The proof for the variable $k^0$ is similar.

\section{The Convolution of two massless Wheeler's Propagators}
\setcounter{equation}{0}

The massless Wheeler's propagator $w_0$ is given by:
\begin{equation}
\label{ep6.1}
w_0(k)=\frac {i} {k^2_0-{\rho}^2}
\end{equation}
It can be extended to the complex plane as a tempered ultradistribution
in the variables $k^0$ and $\rho$:
\[W_0(k)=-i\frac {Sgn\Im(k^0)} {8k^0}
\left[\frac {Sgn\Im(\rho)-Sgn\Im(k^0)}
{\rho -k^0}-\right.\]
\begin{equation}
\label{ep6.2}
\left.\frac {Sgn\Im(\rho)+Sgn\Im(k^0)} {\rho +k^0} \right]
\end{equation}
where $Sgn(x)$ is the function sign of the variable $x$.

We can now evaluate the convolution of two massless
Wheeler's propagators. Then according to (\ref{ep5.18}) and (\ref{ep6.2})
we can write:
\[L_{{\lambda}_0\lambda}(k^0,\rho)=
-\oint\limits_{{\Gamma}^0_1}
\oint\limits_{{\Gamma}^0_2}\oint\limits_{{\Gamma}_1}
\oint\limits_{{\Gamma}_2}
\frac {Sgn\Im(k^0_1)} {8k^0_1}\left[\frac {Sgn\Im({\rho}_1)-Sgn\Im(k^0_1)}
{{\rho}_1 -k^0_1}-
\frac {Sgn\Im({\rho}_1)+Sgn\Im(k^0_1)} {{\rho}_1 +k^0_1} \right]\]
\[\frac {Sgn\Im(k^0_2)} {8k^0_2}\left[\frac {Sgn\Im({\rho}_2)-Sgn\Im(k^0_2)}
{{\rho}_2 -k^0_2}-
\frac {Sgn\Im({\rho}_2)+Sgn\Im(k^0_2)} {{\rho}_2 +k^0_2} \right]\times\]
\begin{equation}
\label{ep6.3}
\frac {k^{0\;{\lambda}_0}_1{\rho}^{{\lambda}+1}_1
k^{0\;{\lambda}_0}_2{\rho}^{{\lambda}+1}_2}
{(k^0-k^0_1-k^0_2)
[{\rho}^2-({\rho}_1+{\rho}_2)^2]}\;
d{\rho}_1\;d{\rho}_2\;dk^0_1\;dk^0_2
\end{equation}
equation (\ref{ep6.3}) can be written as:
\[L_{{\lambda}_0\lambda}(k^0,\rho)=
-\oint\limits_{{\Gamma}^0_1}
\oint\limits_{{\Gamma}^0_2}\; \iint\limits^{+\infty}_{-\infty}
\left\{\frac {Sgn\Im(k^0_1)} {8{\rho}_1}\left[\frac {1} {k^0_1-{\rho}_1}-
\frac {1} {k^0_1+{\rho}_1} \right]\right. \times\]
\[\left[\left({\rho}_1+i0\right)^{\lambda + 1}+
\left({\rho}_1-i0\right)^{\lambda + 1}
\right]+ \frac {1} {8k_1^0}\left[\frac {1} {k^0_1+{\rho}_1}-\frac {1}
{k^0_1-{\rho}_1}\right]\times\]
\[\left.\left[\left({\rho}_1+i0\right)^{\lambda +1}-
\left({\rho}_1-i0\right)^{\lambda + 1}
\right]\right\}\
\left\{\frac {Sgn\Im(k^0_2)} {8{\rho}_2}\left[\frac {1} {k^0_2-{\rho}_2}-
\frac {1} {k^0_2+{\rho}_2} \right]\right.\times\]
\[\left[\left({\rho}_2+i0\right)^{\lambda+1}+
\left({\rho}_2-i0\right)^{\lambda +1}\right]
+\frac {1} {8k_2^0} \left[\frac {1} {k^0_2+{\rho}_2}-\frac {1} {k^0_2-
{\rho}_2} \right]\times\]
\begin{equation}
\label{ep6.4}
\left.\left[\left({\rho}_2+i0\right)^{\lambda +1}
-\left({\rho}_2-i0\right)^{\lambda +1}
\right]\right\}
\frac {k^{0\;{\lambda}_0}_1
k^{0\;{\lambda}_0}_2\;
d{\rho}_1\;d{\rho}_2\;dk^0_1\;dk^0_2}
{(k^0-k^0_1-k^0_2)
{[\rho}^2-({\rho}_1+{\rho}_2)^2]}
\end{equation}
Integrating (\ref{ep6.4}) in the variable $k^0_1$ we obtain
\[L_{\lambda}(k^0,\rho)=-
\oint\limits_{{\Gamma}^0_2}\; \iint\limits^{+\infty}_{-\infty}
\left\{\frac {i\pi} {4{\rho}_1} Sgn\Im(k^0)
\left[\frac {1} {k^0_2-(k^0-{\rho}_1)}-\frac {1} {k^0_2-(k^0+{\rho}_1}
\right]\right. \times\]
\[\left[\left({\rho}_1+i0\right)^{\lambda + 1}+
\left({\rho}_1-i0\right)^{\lambda + 1}
\right]+ \frac {i\pi} {4{\rho}_1}\left[\frac {2} {k^0_2-k^0}-\frac {1}
{k^0_2-(k^0-{\rho}_1)}-\frac {1} {k^0_2-(k^0-{\rho}_1)}\right]\times\]
\[\left.\left[\left({\rho}_1+i0\right)^{\lambda +1}-
\left({\rho}_1-i0\right)^{\lambda + 1}
\right]\right\}\
\left\{\frac {Sgn\Im(k^0_2)} {8{\rho}_2}\left[\frac {1} {k^0_2-{\rho}_2}-
\frac {1} {k^0_2+{\rho}_2} \right]\right.\times\]
\[\left[\left({\rho}_2+i0\right)^{\lambda+1}+
\left({\rho}_2-i0\right)^{\lambda +1}\right]
+\frac {1} {8k_2^0} \left[\frac {1} {k^0_2+{\rho}_2}-\frac {1} {k^0_2-
{\rho}_2} \right]\times\]
\begin{equation}
\label{ep6.5}
\left.\left[\left({\rho}_2+i0\right)^{\lambda +1}
-\left({\rho}_2-i0\right)^{\lambda +1}
\right]\right\}
\frac {d{\rho}_1\;d{\rho}_2\;dk^0_2}
{{\rho}^2-({\rho}_1+{\rho}_2)^2}\;
\end{equation}
where we have selected ${\lambda}_0=0$ due to the fact  the integral is
convergent
for ${\lambda}_0=0$.

There have a sole term in (\ref{ep6.5}) whose integral is not null. It is:
\[L_{\lambda}(k^0,\rho)=-
\oint\limits_{{\Gamma}^0_2}\; \iint\limits^{+\infty}_{-\infty}
\frac {i\pi} {4{\rho}_1} Sgn\Im(k^0)
\left[\frac {1} {k^0_2-(k^0-{\rho}_1)}-\frac {1} {k^0_2-(k^0+{\rho}_1}
\right]\times\]
\[\left[\left({\rho}_1+i0\right)^{\lambda + 1}+
\left({\rho}_1-i0\right)^{\lambda + 1}
\right]
\frac {Sgn\Im(k^0_2)} {8{\rho}_2}\left[\frac {1} {k^0_2-{\rho}_2}-
\frac {1} {k^0_2+{\rho}_2} \right]\times\]
\begin{equation}
\label{ep6.6}
\left[\left({\rho}_2+i0\right)^{\lambda+1}+
\left({\rho}_2-i0\right)^{\lambda +1}\right]
\frac {d{\rho}_1\;d{\rho}_2\;dk^0_2}
{{\rho}^2-({\rho}_1+{\rho}_2)^2}\;
\end{equation}

Evaluation of (\ref{ep6.6}) gives:
\[L_{\lambda}(k^0,\rho)=\frac {{\pi}^2k^0} {2}
\iint\limits_{-\infty}^{+\infty}
\left[({\rho}_1+ i0)^{\lambda + 1}+({\rho}_1-i0)^{\lambda +1}\right]
 \left[({\rho}_2+ i0)^{\lambda + 1}+({\rho}_2-i0)^{\lambda +1}\right]\]
\begin{equation}
\label{ep6.7}
\frac {d{\rho}_1\;d{\rho}_2}
{\left[(k_0^2+{{\rho}_1}^2-{{\rho}_2}^2)^2-4k_0^2{{\rho}_1}^2\right]
\left[{\rho}^2-({\rho}_1+{\rho}_2)^2\right]}
\end{equation}
We can evaluate now the integral in the variable
${\rho}_2$ in (\ref{ep6.7}). The result is:
\[L_{\lambda}(k^0,\rho)=
\frac {{\pi}^3} {16\rho}\frac {(1+\cos\pi\lambda)^2}
{\sin\frac {\pi(\lambda+1)}{2}}\int\limits_0^{\infty}\;d{\rho}_1\;\;
{\rho}_1^{\lambda}\;\times\]
\[\left\{\frac {e^{-\frac {i\pi} {2}(\lambda+1)
Sgn\Im(k^0)}(k^0+{\rho}_1)^{
\lambda+1}-
e^{-\frac {i\pi} {2}(\lambda+1)Sgn\Im(\rho)}(\rho+{\rho}_1)^{
\lambda+1}}{(\rho-k^0)\left(\frac {\rho+k^0} {2}+
{\rho}_1\right)}\;-\right.\]
\[\frac {e^{-\frac {i\pi} {2}(\lambda+1)Sgn\Im(k^0)}(k^0+{\rho}_1)^{
\lambda+1}-
e^{\frac {i\pi} {2}(\lambda+1)Sgn\Im(\rho)}({\rho}_1-\rho)^{
\lambda+1}}{(\rho+k^0)\left(\frac {\rho-k^0} {2}-{\rho}_1\right)}\;-\]
\[\frac {e^{\frac {i\pi} {2}(\lambda+1)Sgn\Im(k^0)}({\rho}_1-k^0)^{
\lambda+1}-
e^{-\frac {i\pi} {2}(\lambda+1)Sgn\Im(\rho)}({\rho}_1+\rho)^{
\lambda+1}}{(\rho+k^0)\left(\frac {\rho-k^0} {2}+{\rho}_1\right)}\;+\]
\begin{equation}
\label{ep6.8}
\left.\frac {e^{\frac {i\pi} {2}(\lambda+1)Sgn\Im(k^0)}({\rho}_1-k^0)^{
\lambda+1}-
e^{\frac {i\pi} {2}(\lambda+1)Sgn\Im(\rho)}({\rho}_1-\rho)^{
\lambda+1}}{(\rho-k^0)\left(\frac {\rho+k^0} {2}-{\rho}_1\right)}\right\}
\end{equation}
The evaluation of  (\ref{ep6.8}) is tedious task. Fortunately
$\lim\; \lambda\rightarrow 0$ can be taken without problem
in the finals steps of
the calculation. The result is:
\[L(k^0,\rho)=\frac {{\pi}^3} {4\rho}\left[
\frac {\pi} {2} Sgn\Im(k^0)Sgn\Im(k^0+\rho) + \frac {\pi} {2}
Sgn\Im(\rho)Sgn\Im(k^0+\rho) \right.\;+ \]
\begin{equation}
\label{ep6.9}
\left.
\frac {\pi} {2} Sgn\Im(k^0)Sgn\Im(\rho-k^0)-
Sgn\Im(\rho-k^0)\right]
\end{equation}
Eq. (\ref{ep6.9}) can be written:
\[L(k^0,\rho)=\frac {{\pi}^4} {8\rho} \left[\left(Sgn\Im(k^0)+
Sgn\Im(\rho)\right)
Sgn\Im(\rho+k^0)\right.\; +\]
\[\left. \left(Sgn\Im(k^0)-Sgn\Im(\rho)\right)Sgn\Im(\rho-k^0)\right]\,=\]
\begin{equation}
\label{ep6.10}
\frac {{\pi}^4} {4\rho} \; Sgn\Im(k^0)Sgn\Im(\rho)
\end{equation}
Taking into account that:
\[H(k^0,\rho)=\frac {1} {2\pi\rho} \int L(k^0,\rho) \rho\;d\rho\]
we obtain:
\begin{equation}
\label{ep6.11}
H(k^0,\rho)=\frac {{\pi}^3} {8} Sgn\Im(k^0)Sgn\Im(\rho)=
[W_0\ast W_0](k^0, \rho)
\end{equation}
(The symbol $\ast$ indicates  the convolution product).

Thus the cut of $H(k^0, \rho)$ along the real axis, i.e., the distribution
$h(k^0, \rho)$ is:
\begin{equation}
\label{ep6.12}
h(k^0, \rho)=\frac {{\pi}^3} {2}=[w_0\ast w_0](k^0, \rho)
\end{equation}

\section{The Convolution of two complex mass Wheeler's Propagators}
\setcounter{equation}{0}

The complex mass Wheeler's propagator is:
\begin{equation}
\label{ep7.1}
w_{\mu}(x)=-\frac {i\pi} {2} \frac {{\mu}^{n/2-1}} {(2\pi)^{n/2}}
Q^{1/2(1-n/2)}_- J_{1-n/2}(\mu Q_-^{1/2})
\end{equation}
and it Fourier transform has the expression:
\[W_{\mu}(k^0, \rho)=-\frac {iSgn[\Im(k^0)]} {8\sqrt{k_0^2-{\mu}^2}}
\left[\frac {Sgn[\Im(\rho)]-Sgn[\Im(\sqrt{k_0^2-{\mu}^2})]}
{\rho - \sqrt{k_0^2-{\mu}^2}}\;-\right. \]
\begin{equation}
\label{ep7.2}
\left.\frac {Sgn[\Im(\rho)]+Sgn[\Im(\sqrt{k_0^2-{\mu}^2})]}
{\rho + \sqrt{k_0^2-{\mu}^2}}\right]
\end{equation}
Using (\ref{ep7.2}) we have now:
\[L(k^0,\rho)=-\oint\limits_{{\Gamma}^0_1}
\oint\limits_{{\Gamma}^0_2}\oint\limits_{{\Gamma}_1}
\oint\limits_{{\Gamma}_2}
\frac {Sgn[\Im(k^0_1)]} {8\sqrt{k_1^{02}-{\mu}_1^2}}
\left[\frac {Sgn[\Im({\rho}_1)]-Sgn[\Im(\sqrt{k_1^{02}-{\mu}_1^2})]}
{{\rho}_1 - \sqrt{k_1^{02}-{\mu}_1^2}}\;-\right. \]
\[\left.\frac {Sgn[\Im({\rho}_1)]+Sgn[\Im(\sqrt{k_1^{02}-{\mu}_1^2})]}
{{\rho}+1 + \sqrt{k_1^{02}-{\mu}_1^2}}\right]
\frac {Sgn[\Im(k^0_2)]} {8\sqrt{k_2^{02}-{\mu}_2^2}}
\left[\frac {Sgn[\Im({\rho}_2)]-Sgn[\Im(\sqrt{k_2^{02}-{\mu}_2^2})]}
{{\rho}_2 - \sqrt{k_2^{02}-{\mu}_2^2}}\;-\right. \]
\begin{equation}
\label{ep7.3}
\left.\frac {Sgn[\Im({\rho}_2)]+Sgn[\Im(\sqrt{k_2^{02}-{\mu}_2^2})]}
{{\rho}_2 + \sqrt{k_2^{02}-{\mu}_2^2}}\right]
\frac {{\rho}_1{\rho}_2d{\rho}_1\;d{\rho}_2\;dk_1^0\;dk^0_2} {(k^0-
k^0_1-k^0_2)[{\rho}^2-({\rho}_1+{\rho}_2)^2]}
\end{equation}
where we have selected ${\lambda}_0=\lambda=0$ due to that (\ref{ep7.3})
is convergent in this point (Observe the reader that it is due to the definition of
$L(k^0,\rho)$). Now (\ref{ep7.3}) is equal to:
\[L(k^0,\rho)=-\frac {1} {4} \oint\limits_{{\Gamma}^0_1}
\oint\limits_{{\Gamma}^0_2}\;
\iint\limits_{-\infty}^{+\infty}
\frac {Sgn[\Im(k^0_1)]} {{\rho}_1^2+{\mu}_1^2-k_1^{02}}\;\;
\frac {Sgn[\Im(k^0_2)]} {{\rho}_2^2+{\mu}_2^2-k_2^{02}}\;\times\]
\begin{equation}
\label{ep7.4}
\frac {{\rho}_1{\rho}_2} {(k^0-
k^0_1-k^0_2)[{\rho}^2-({\rho}_1+{\rho}_2)^2]}
d{\rho}_1\;d{\rho}_2\;dk^0_1\;dk^0_2
\end{equation}
and can be re-written as:
\[L(k^0,\rho)=-\frac {1} {16} \oint\limits_{{\Gamma}^0_1}
\oint\limits_{{\Gamma}^0_2}\;
\iint\limits_{-\infty}^{+\infty}
\frac {Sgn[\Im(k^0_1)]} {\sqrt{{\rho}_1^2+{\mu}_1^2}}
\left[\frac {1} {k^0_1-\sqrt{{\rho}_1^2+{\mu}_1^2}}-
\frac {1} {k^0_1+\sqrt{{\rho}_1^2+{\mu}_1^2}}\right]\;\times\]
\[\frac {Sgn[\Im(k^0_2)]} {\sqrt{{\rho}_2^2+{\mu}_2^2}}
\left[\frac {1} {k^0_2-\sqrt{{\rho}_2^2+{\mu}_2^2}}-
\frac {1} {k^0_2+\sqrt{{\rho}_2^2+{\mu}_2^2}}\right]
\frac {1} {(k^0-k^0_1-k^0_2)}\times\]
\begin{equation}
\label{ep7.5}
\frac {{\rho}_1{\rho}_2} {{\rho}^2-({\rho}_1+{\rho}_2)^2}
d{\rho}_1\;d{\rho}_2\;dk^0_1\;dk^0_2
\end{equation}
Taking into account that:
\[\oint\limits_{{\Gamma}^0_1}
\oint\limits_{{\Gamma}^0_2}
\frac {Sgn[\Im(k^0_1)]Sgn[\Im(k^0_2)} {k^0-k^0_1-k^0_2}
\left[\frac {1} {k^0_1-\sqrt{{\rho}_1^2+{\mu}_1^2}}-
\frac {1} {k^0_1+\sqrt{{\rho}_1^2+{\mu}_1^2}}\right]\;\times\]
\[\left[\frac {1} {k^0_2-\sqrt{{\rho}_2^2+{\mu}_2^2}}-
\frac {1} {k^0_2+\sqrt{{\rho}_2^2+{\mu}_2^2}}\right]\;dk^0_1\;dk^0_2=\]
\begin{equation}
\label{ep7.6}
-\frac {32{\pi}^2k^0\sqrt{{\rho}_1^2+{\mu}_1^2}
\sqrt{{\rho}_2^2+{\mu}_2^2}} {[k_0^2+({\rho}_2^2+{\mu}_2^2)-
({\rho}_1^2+{\mu}_2^2)]^2-4k_0^2({\rho}_2^2+{\mu}_2^2)}
\end{equation}
Replacing this result in (\ref{ep7.5}) we obtain
\[L(k^0,\rho)=2{\pi}^2k^0
\iint\limits_{-\infty}^{+\infty}\frac {1}
{[k_0^2+({\rho}_2^2+{\mu}_2^2)-
({\rho}_1^2+{\mu}_2^2)]^2-4k_0^2({\rho}_2^2+{\mu}_2^2)}\;\times\]
\begin{equation}
\label{ep7.7}
\frac {{\rho}_1{\rho}_2} {{\rho}^2-({\rho}_1+{\rho}_2)^2}
\;d{\rho}_1\;d{\rho}_2
\end{equation}
Taking into account that
\begin{equation}
\label{ep7.8}
\int\frac {\rho\;d\rho} {{\rho}^2-({\rho}_1+{\rho_2})^2}=
\Theta[\Im(\rho)]\ln({\rho}_1+{\rho}_2-\rho)+\Theta[-\Im(\rho)]
\ln(\rho-{\rho}_1-{\rho}_2)
\end{equation}
and using the result (\ref{ep7.7}) we obtain
\[H(k^0,\rho)=\frac {\pi k^0} {\rho}
\iint\limits_{-\infty}^{+\infty}\frac {1}
{[k_0^2+({\rho}_2^2+{\mu}_2^2)-
({\rho}_1^2+{\mu}_2^2)]^2-4k_0^2({\rho}_2^2+{\mu}_2^2)}\;\times\]
\begin{equation}
\label{ep7.9}
\Theta[\Im(\rho)]\ln({\rho}_1+{\rho}_2-\rho)+\Theta[-\Im(\rho)]
\ln(\rho-{\rho}_1-{\rho}_2)\;d{\rho}_1\;d{\rho}_2
\end{equation}
The equation (\ref{ep7.9}) can be written in the real $\rho$-axis as:
\begin{equation}
\label{ep7.10}
H(k^0,\rho)=\frac {i{\pi}^2  k^0} {\rho}
\iint\limits_{-\infty}^{+\infty}
\frac {Sgn({\rho}_1+{\rho}_2-\rho)\;{\rho}_1{\rho}_2\;d{\rho}_1\;d{\rho}_2}
{[k_0^2+({\rho}_2^2+{\mu}_2^2)-
({\rho}_1^2+{\mu}_2^2)]^2-4k_0^2({\rho}_2^2+{\mu}_2^2)}
\end{equation}
After the evaluation of double integral of  (\ref{ep7.10}) we obtain:
\[H(k^0,\rho)=\frac {{\pi}^3Sgn[\Im(k^0)]} {4(k_0^2-{\rho}^2)}
\sqrt{(k_0^2-{\rho}^2+{\mu}_2^2-{\mu}_1^2)^2-4(k_0^2-{\rho}^2){\mu}_2^2}=\]
\begin{equation}
\label{ep7.11}
[W_{{\mu}_1}\ast W_{{\mu}_2}](k^0,\rho)
\end{equation}

\section{Discussion}

In a earlier paper \cite{tp3} we have shown the existence of the convolution
of two one-dimensional tempered ultradistributions. In this paper we have
extended these procedure to n-dimensional space. In four-dimensional
space we have obtained a expression for the convolution of two tempered
ultradistributions even in the variables $k^0$ and $\rho$.

When we use the perturbative development in Quantum Field Theory, we
have to deal with products of distributions in configuration space,
or else, with convolutions in the Fourier transformed p-space.
Unfortunately, products or convolutions ( of distributions ) are
in general ill-defined quantities. However, in physical applications
one introduces some ``regularization'' scheme, which allows us to
give sense to divergent integrals. Among these procedures we would
like to mention the dimensional regularization method ( ref.
\cite{tp14,tp15} ). Essentially, the method consists in the
separation of the volume element ( $d^{\nu}p$ ) into an angular
factor ( $d\Omega$ ) and a radial factor ( $p^{\nu-1}dp$ ).
First the angular integration is carried out and then the number
of dimensions $\nu$ is taken as a free parameter. It can be adjusted
to give a convergent integral, which is an analytic function of
$\nu$.

Our formula (4.1) is similar to the expression one obtains with
dimensional regularization. However, the parameters $\lambda$ are
completely independents of any dimensional interpretation.

All ultradistributions provide integrands ( in (4.1) ) that are
analytic functions along the integration paths. The parameters
$\lambda$ permit us to control the possible tempered asymptotic
behavior ( cf. eq. (3.9) ). The existence of a region of
analyticity for each $\lambda$, and a subsequent continuation to
the point of interest ( ref. \cite{tp3} ), defines the convolution
product.

For tempered ultradistributions (even in the variables $k^0$ and $\rho$)
we have  obtained formula (\ref{ep5.15}) for which are valid similar
considerations to those
given for (\ref{ep4.1})
The properties described below
show that tempered ultradistributions provide an
appropriate framework for applications to physics. Furthermore,
they can ``absorb'' arbitrary pseudo-polynomials, thanks to eq. (3.10).
A property that is interesting for renormalization theory.
 For this reason we decided to begin this paper and also for the benefit
 of the reader
we began this paper with a summary of the main characteristics
of n-dimensional tempered ultradistributions and their Fourier
transformed
distributions of the exponential type.

\newpage

\end{document}